\documentclass{aa}
\usepackage{latexsym}
\usepackage{epsfig}
\usepackage{times}
\usepackage{graphicx}
\def\kms{km~s$^{-1}$}
\def\0{\hspace*{0.5em}}

\begin{document}
\title{The radial velocity curve of HD153919 (4U1700-37) revisited\thanks{Based on observations obtained with the {\it International Ultraviolet Explorer}}}

\author{G. Hammerschlag-Hensberge\inst{1} \and M.H. van Kerkwijk\inst{2,3} \and L. Kaper\inst{1}}

\offprints{G. Hammerschlag-Hensberge;  e-mail: godeliev@science.uva.nl}

\institute{Astronomical Institute ``Anton Pannekoek'' and Center for
           High-Energy Astrophysics, 
           University of Amsterdam, Kruislaan 403, 1098 SJ Amsterdam, 
           The Netherlands
           \and Astronomical Institute, Utrecht University, P.O. Box
80000, 3508 TA Utrecht, The Netherlands
\and Dept. of Astronomy \& Astrophysics, 60 St George Street, Toronto,
ON, M5S 3H8, Canada}

\date{Received; Accepted}

\authorrunning{Hammerschlag-Hensberge et al.}
\titlerunning{The radial velocity of HD153919/4U1700-37}

\abstract{We have re-analysed all available high-resolution
ultraviolet IUE spectra of the high-mass X-ray binary
HD153919/4U1700-37. The radial velocity semi-amplitude of $20.6 \pm
1.0$~\kms\ and orbital eccentricity of $0.22 \pm 0.04$ agree very well
with the values obtained earlier from optical spectra. They disagree
with earlier conclusions for the same data reduced by Heap \& Corcoran
(\cite{HC92}) and by
Stickland \& Lloyd (\cite{SL93}).
}

\maketitle

\begin{table*}
\caption[]{Journal of {\it IUE} SWP high-resolution spectra of
HD153919. Image numbers include an S or an L, refering to the small and
large aperture, respectively. Phases are calculated from the ephemeris
 $T_{0}=JD2448900.873 \pm 0.015$; $P=3.411581 \pm 0.000027$ given by Rubin et
al. (\cite{RF96}) including the orbital period change  $\dot{P}/P= -(3.3 \pm
0.6) \times 10^{-6}$~yr$^{-1}$. The velocities
are corrected for a shift of interstellar lines and are relative to that of
SWP 1476 which was arbitrarily set to zero.}
\begin{flushleft}
\begin{tabular}{lllrrlllrr} \hline
\multicolumn{10}{c}{Journal of observations} \\
\hline \hline
SWP & $JD_{\rm mid.exp.}$  &     Phase & $v_{\rm rad}$  &
$1\sigma$-error & \hspace{1cm} SWP & $JD_{\rm mid.exp.}$ & Phase & $v_{\rm rad}$ & $1\sigma$-error\\
& (-244000) & & (\kms) & (\kms) & & (-244000) && (\kms) & (\kms)\\
\hline
  1476S & 3632.610 & 0.8076 &   0.0 & 2.9  &\hspace{1cm}   25586L & 6160.579 & 0.7771 &  -0.7 & 2.8 \\
  1513S & 3638.414 & 0.5087 &  -3.2 & 5.9  &\hspace{1cm}   25587L & 6160.618 & 0.7885 &  -5.7 & 2.9\\
  1714S & 3664.042 & 0.0204 &  15.7 & 2.7  &\hspace{1cm}   25588L & 6160.665 & 0.8023 &   0.5 & 2.4\\ 
  1960L & 3700.269 & 0.6388 &  -6.4 & 4.5  &\hspace{1cm}   25589L & 6160.715 & 0.8169 &   2.4 & 2.5\\
  1961L & 3700.308 & 0.6502 &  -6.5 & 4.3  &\hspace{1cm}   25590L & 6160.760 & 0.8300 &  -0.4 & 2.6\\
  1969L & 3701.030 & 0.8618 &   1.9 & 3.4  &\hspace{1cm}   25591L & 6160.802 & 0.8424 &  -1.1 & 2.7\\
  1970L & 3701.085 & 0.8779 &   1.9 & 2.7  &\hspace{1cm}   25592L & 6160.844 & 0.8548 &   2.8 & 2.6\\
  1972S & 3701.231 & 0.9207 &   5.5 & 3.8  &\hspace{1cm}   25596L & 6161.673 & 0.0977 &  21.1 & 2.4\\
  1973S & 3701.280 & 0.9351 &   9.7 & 7.6  &\hspace{1cm}   25597L & 6161.706 & 0.1074 &  26.8 & 2.2\\
  1975S & 3701.409 & 0.9729 &   4.6 & 2.9  &\hspace{1cm}   25598L & 6161.759 & 0.1229 &  22.6 & 2.6\\
  1982S & 3702.224 & 0.2118 &  22.2 & 5.0  &\hspace{1cm}   25599L & 6161.808 & 0.1373 &  22.8 & 2.2\\
  1983S & 3702.272 & 0.2258 &  23.6 & 3.3  &\hspace{1cm}   25600L & 6161.839 & 0.1463 &  25.8 & 2.2\\
  1986S & 3702.404 & 0.2645 &  24.4 & 4.3  &\hspace{1cm}   25607L & 6162.582 & 0.3642 &  29.2 & 2.1\\
  1987S & 3702.474 & 0.2851 &  36.0 & 2.1  &\hspace{1cm}   25608L & 6162.618 & 0.3748 &  29.3 & 2.3\\
  1991S & 3703.099 & 0.4683 &   8.8 & 3.0  &\hspace{1cm}   25609L & 6162.651 & 0.3844 &  25.6 & 2.2\\
  1992S & 3703.148 & 0.4826 &   6.2 & 3.4  &\hspace{1cm}   25610L & 6162.689 & 0.3955 &  27.6 & 2.5\\
  1994S & 3703.258 & 0.5148 &   5.4 & 2.9  &\hspace{1cm}   25611L & 6162.728 & 0.4070 &  25.5 & 2.3\\
  1995S & 3703.308 & 0.5295 &   5.8 & 3.0  &\hspace{1cm}   25612L & 6162.764 & 0.4176 &  23.5 & 2.3\\
  2002S & 3703.617 & 0.6200 &  -4.7 & 2.1  &\hspace{1cm}   25613L & 6162.800 & 0.4280 &  21.8 & 2.2\\
  2003S & 3703.708 & 0.6467 &  -1.7 & 2.0  &\hspace{1cm}   25614L & 6162.833 & 0.4377 &  28.7 & 2.3\\
  2004S & 3703.770 & 0.6649 &  -0.7 & 2.1  &\hspace{1cm}   25615L & 6162.892 & 0.4551 &  17.0 & 2.1\\
  2006S & 3704.067 & 0.7519 &  -6.2 & 2.8  &\hspace{1cm}   25616L & 6162.929 & 0.4659 &  16.4 & 1.9\\
  2008S & 3704.156 & 0.7780 & -12.9 & 3.0  &\hspace{1cm}   25617L & 6162.970 & 0.4780 &  12.0 & 1.8\\
  2009S & 3704.206 & 0.7927 &  -8.2 & 2.6  &\hspace{1cm}   25618L & 6163.005 & 0.4881 &   8.1 & 2.1\\
  2106S & 3715.317 & 0.0494 &  17.6 & 2.5  &\hspace{1cm}   25619L & 6163.044 & 0.4996 &   5.7 & 5.4\\
  2153S & 3720.156 & 0.4677 &   9.2 & 2.3  &\hspace{1cm}   25620L & 6163.147 & 0.5298 &   0.1 & 2.3\\
  4742S & 3957.201 & 0.9470 &   6.6 & 2.2  &\hspace{1cm}   25621L & 6163.187 & 0.5415 &  -5.4 & 2.3\\
  4751S & 3958.040 & 0.1929 &  34.6 & 2.2  &\hspace{1cm}   28730L & 6632.663 & 0.1509 &  31.3 & 2.5\\
  4752S & 3958.088 & 0.2070 &  35.9 & 2.0  &\hspace{1cm}   28731L & 6632.707 & 0.1637 &  34.5 & 2.3\\
  4753S & 3958.138 & 0.2216 &  36.4 & 2.5  &\hspace{1cm}   28732L & 6632.753 & 0.1772 &  34.3 & 2.4\\
  5180S & 4003.047 & 0.3848 &  24.7 & 2.7  &\hspace{1cm}   & & & & \\
\hline					
\end{tabular}
\end{flushleft}
\label{table1}
\end{table*}

\section{Introduction}

The bright Of star \object{HD153919} is the optical counterpart of the
3.4 day eclipsing high-mass X-ray binary \object{4U1700-37}, discovered
by Jones et al. (\cite{Jo73}). About 2.5 million years ago, the binary
system escaped from the OB association Sco OB2 after the supernova
explosion that formed the compact companion (Ankay et al. \cite{AK01}).  The X-ray
source powered by the stellar wind of the Of star shows no regular
pulsations and the highly variable X-ray flux is less
than expected from standard wind accretion (e.g. Rubin et
al. \cite{RF96}, Kaper \cite{Ka98}). The decrease in orbital period
determined from mid-eclipse observations and first mentioned by Haberl
et al. (\cite{HW89}) was confirmed by Rubin et al. (\cite{RF96}) and
is probably due to wind-driven angular momentum loss from the
Of star.  Because no direct measurements can be made of the
orbital motion of the compact object, system parameters depend heavily
on radial velocity studies of the optical companion. The most recent
analysis of the physical parameters cannot resolve the nature of the
compact object in this system (Clark et al. \cite{CG02}). Its likely
mass is considerably lower than those found for black holes, but it is
significantly in excess to fit the high density nuclear equation of
state for neutron stars. A point to remember, however, is that also
the Vela X-1 pulsar has a high mass, comparable to that of 4U1700-37
(van Kerkwijk et al. \cite{vK95}).

The most thorough study of radial velocities of lines in the visible
spectral region has been made by Hammerschlag-Hensberge
(\cite{HH78}). It is clear from the observations that the strongly
variable stellar wind in this system complicates a straightforward
determination of the system parameters. The results for the orbital
analysis of all lines together in the visible blue spectral region
($3700-4900$~ \AA) yield a semi-amplitude $K = 19 \pm 1$~\kms\ and an
orbital eccentricity $e = 0.16 \pm 0.08$.

Heap \& Corcoran (\cite{HC92}) have examined the system parameters
using the available ultraviolet spectra of HD153919. They find $K=18
\pm 3$~\kms\ and a negligible eccentricity. For their investigation
they cross-correlated a few selected narrow regions of the ultraviolet
spectra with the same regions of one of the spectra.  To bring their
spectra to the same frame of reference they used the heliocentric
correction. However, instrumental and processing shifts during the
reduction of spectra obtained with the {\it International Ultraviolet
Explorer} (IUE) require a more sophisticated approach (Stickland
\cite{St92}).

Stickland \& Lloyd (\cite{SL93}) re-examined the same data, now
cross-correlating both stellar and interstellar spectra with reference
to one of the UV spectra. They used the complete ultraviolet spectral
range excluding P-Cygni lines whose profiles are affected by mass
loss. Amazingly, they found a much smaller semi-amplitude $K$ of
10~\kms\ and an eccentricity of 0.17.

In a similar study of the high-mass X-ray binary HD77581/Vela X-1
Barziv et al. (\cite{BK00}) found a disagreement in the semi-amplitude
of the radial velocity curve with the results of Stickland et
al. (\cite{SL97}) for this system.  This strengthened our decision to
re-analyse the ultraviolet spectra of HD153919 again. In Sect. 2 we
describe the observations and the data reduction procedures. In
Sect. 3 we present the resulting radial velocity curve and in
Sect. 4 we compare our results with the ones found by Heap \&
Corcoran and by Stickland \& Lloyd. In Sect. 5 we summarize our
conclusions and discuss the evolutionary consequences.

\section{Observations and data reduction}

A large number of ultraviolet spectra of HD153919 have been obtained
with the {\it International Ultraviolet Explorer}. The spectra
obtained with the Short Wavelength Prime Camera in high-resolution
mode are listed in Table~\ref{table1}. A separate analysis of the same
IUE spectra was performed by Stickland \& Lloyd (\cite{SL93}). These
authors cross-correlated the set of spectra with one particular
spectrum (SWP25600). Because their results were in large discrepancy
with results from optical data we decided to re-analyse the same IUE
data.  We cross-correlated each spectrum with {\it all}
other spectra of the data set to reduce errors. We did not use a template spectrum of another star
with identical spectral type as independent reference spectrum- this is not realistic for such hot
mass-losing stars which each have unique spectra- nor did we use an
average spectrum to avoid auto-correlation effects leading to a
systematic decrease of the amplitude of the radial velocity curve.

First the spectra - initially mapped onto a uniform wavelength grid of
$0.05$~\AA\ - were transformed to a logarithmic wavelength scale
such that the Doppler shift becomes a linear displacement along the
spectrum. The relevant
part of the cross-correlation function is the area close to the top;
it is this upper part which is fitted
with an analytic function (a Gaussian plus a linear function) whose
maximum determines the velocity shift of the spectrum. Because the
cross-correlation profiles of IUE spectra show small excess peaks (see
Evans (\cite{Ev88}) and Fig. 4 in Van Kerkwijk et
al. (\cite{VK95})) due to the so-called fixed-pattern noise in the
detector, we corrected for those by including an extra Gaussian peak
in the fit. For details on the cross-correlation method and instructive
figures we refer to Van Kerkwijk (\cite{VK93}).

The spectral regions which were used for the cross-correlation are
listed in Table~\ref{table2}. The regions were chosen in a way to
exclude parts of the spectrum containing variable line profiles such
as the well-known P-Cygni profiles (see for instance
Hammerschlag-Hensberge et al. (\cite{HHK90})); also the spectral
region including variable emission lines $1540-1740$~\AA\ (Kaper et
al. \cite{KH90}) was excluded. Table~\ref{table2} lists the
wavelength regions including interstellar lines which were used to
place the individual spectra onto the same frame of reference, because
IUE spectra have a variety of instrumental and processing shifts, as
mentioned before.

\begin{table}
\centering
\caption[]{Wavelength regions used for cross-correlation.}
\begin{tabular}{ll} \hline
stellar &\hspace{1cm}  interstellar\\
\hline 
1305-1325 &\hspace{1cm} 1303.5-1305.5\\
1337-1368 &\hspace{1cm} 1333.5-1335.5\\
1425-1490 &\hspace{1cm} 1607.5-1609.5\\
1755-1785 &\hspace{1cm} 1670.0-1672.0\\
1810-1842 &\hspace{1cm} 1807.3-1808.7\\
\hline
\end{tabular}
\label{table2}
\end{table}

\section{Results}

\begin{figure*}
\centering
\includegraphics[width=17cm]{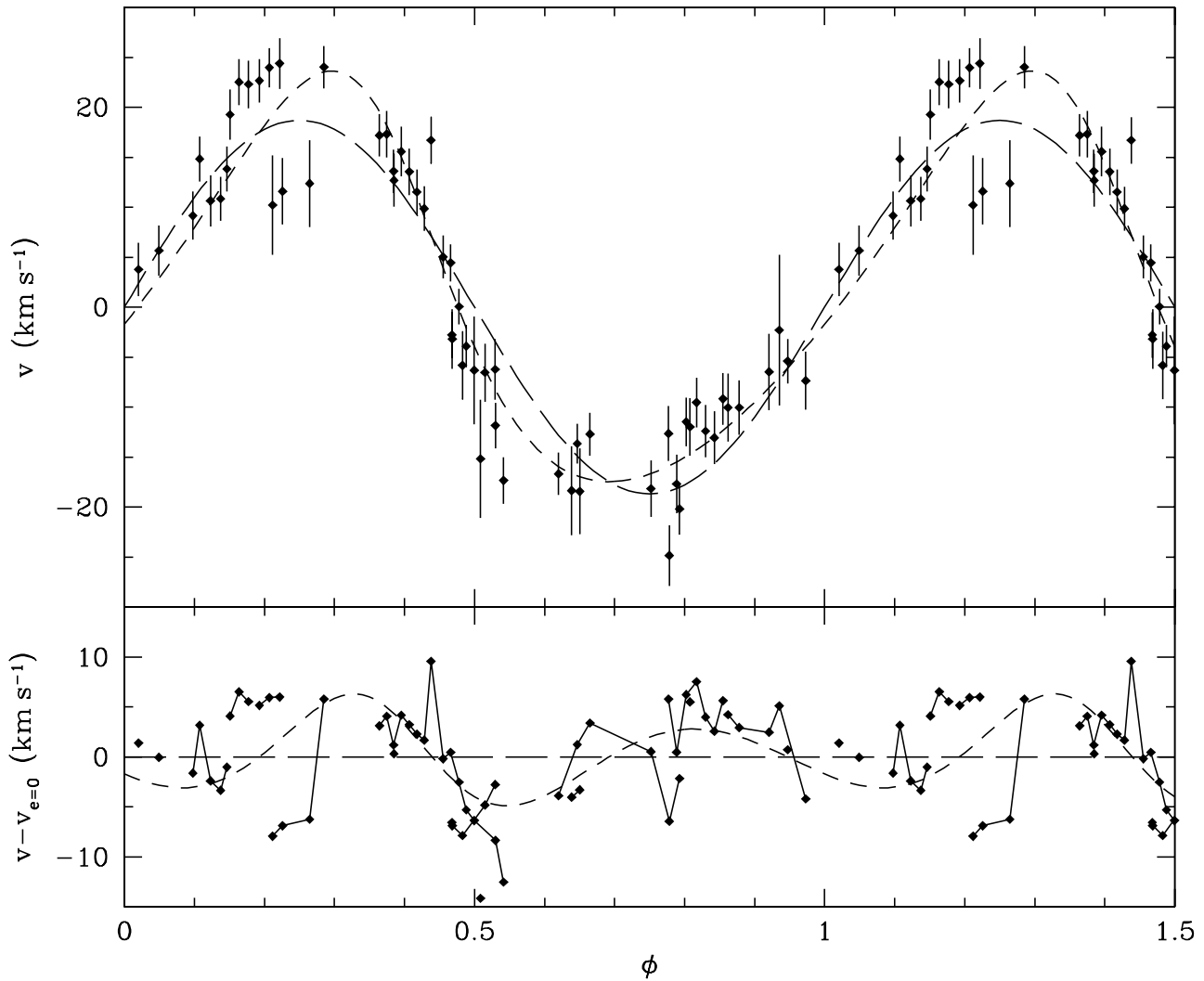}
\caption[]{Radial-velocity curve derived from the cross-correlated IUE
spectra of HD153919. The radial velocities in the top panel are
plotted with error bars indicating the 1$\sigma$
uncertainties. Overdrawn is the Keplerian curve that best fits the
data (short dashes) as well as the solution for a circular orbit (long
dashes). The velocities have been shifted to zero system velocity
based on the best fitting circular orbit (our analysis did not allow
us to derive the systemic velocity; see text). In the lower panel, the
residuals relative to the best-fit circular orbit are shown; the
short-dashed curve shows where the residuals should lie for the
best-fit eccentric orbit. For clarity, the error bars have been
omitted.  Points connected by lines were taken as a sequence.
\label{rvcurve}}
\end{figure*}

In Table~\ref{table1} we list the radial velocities derived from the
cross-correlations of all spectra with each other. In
Fig.~\ref{rvcurve}, we show the radial velocities we obtained from the
IUE spectra as a function of orbital phase. Because the spectra were
cross-correlated with each other, no
 $\gamma$-velocity has been
determined for the system and the velocities in the figure have been
shifted to zero system velocity. The $\gamma$-velocity has been
determined elsewhere (e.g. Gies \& Bolton \cite{GB86}, Humphreys
\cite{Hu78}): $-60$~\kms. It was not included in our figure because
the value - which is rather uncertain - did not follow from our data. At first glance, it is already clear that
the data show significant deviations from a smooth Keplerian curve.
We believe these are intrinsic to the star, since earlier optical
observations have already shown that the star has intrinsic variations
(e.g., Kaper et al.\ \cite{KH94}).  As we will find that the
deviations are substantially larger than the measurement errors, we
give each measurement equal weight in our fits below.

We first fit a circular orbit, shown by the long-dashed line in
Fig.~\ref{rvcurve}, and find $K = 18.7 \pm 1.0~$\kms.  The
root-mean-square (rms) residual is $6.36$~\kms.  This is a factor two
larger than the mean error in the measured velocities (using the
formal errors, one finds a reduced $\chi^{2}=3.6$).  

The residuals relative to the best-fit circular orbit are shown in the
lower panel of Fig.~\ref{rvcurve}.  These seem to show a double-wave
pattern, suggestive of an eccentric orbit.  Fitting an eccentric
orbit, we find eccentricity $e = 0.22 \pm 0.04$, periastron angle
$\omega = 49\degr \pm 11\degr$, and the radial-velocity amplitude is $K =
20.6 \pm 1.0$~\kms.  The rms residuals are slightly lower, at
$5.85$~\kms.  Doing an F-test, this is not a significant improvement.
The F-test, however, ignores the correlations in the residuals.  In
order to obtain a more robust estimate of the significance, we ran
Monte-Carlo simulations, in which we made artificial data sets in
which the velocity at each measurement phase was given by the sum of the
velocity expected from the best-fit circular orbit and a random
deviation.  For the latter, we used two components, one to mimic the
measurement error (i.e., a random number drawn from a normal
distribution with standard deviation equal to the measurement error),
and one to mimic the intrinsic deviations.  For the intrinsic
deviations, we again assumed a normal distribution, with a standard
deviation of 6~\kms, so that our artificial data sets have the same
rms residuals as our data.  Our simulations confirmed our
error estimate on the eccentricity, and showed that the probability of
obtaining an eccentricity of 0.22 by chance due to the deviations was
negligible. 

We should note that what is proven above is that for a circular orbit
there are systematic variations with orbital phase over and above the
significant random excursions.  An eccentric orbit is a simple
explanation for these systematic variations, but we cannot exclude
other systematic effects.  A cause for caution is that Barziv et al.\
(\cite{BK00}) found from their optical data an eccentricity which was
larger than the value from the X-ray pulsar in Vela~X-1 (see their
Fig.~18).  Furthermore, model calculations of Zuiderwijk et al.\
(\cite{vP77}) showed that the radial-velocity curve for a tidally
distorted star in a circular orbit might show an apparent
eccentricity.  The shape of the expected curve, however, is different
from that observed.  Given this, we cannot dismiss the possibility of
a circular orbit out of hand, although we believe that the simplest
and most likely solution is that it is in fact eccentric.

It is obvious that for some spectra the individual radial-velocity
measurements deviate significantly from the best-fit Keplerian
orbit. This was already known from the optical spectra. The Of
supergiant has a strong stellar wind which can cause irregular
variations in the radial velocities of some lines. Also the presence
of a photo-ionization wake (Kaper et al. \cite{KH94}) trailing the
X-ray source may cause irregular small orbit-to-orbit variations in
the radial velocities.  Nevertheless we devoted extra attention to
some of the spectra. Particularly striking are the three velocities
near phase 0.25 that are much lower than the other spectra at the same
phase interval.  These three are for spectra SWP1982, 1983 and 1986,
which were taken in sequence.  This would suggest an instrumental
problem, were it not that SWP1987, which was taken immediately after SWP1986 at the same night as the mentioned three spectra, has a much
higher velocity comparable with that of SWP~4751, 4752 and 4753. We
compared the line profiles of all those spectra but could not find
significant differences, although SWP~1982 and 1983 are very
noisy. Note that the modest spectral resolution ($R \sim 10,000$) and
low S/N ($\sim 20$) does not allow for a detailed study of individual
photospheric lines. 

For HD~77581, the companion of Vela~X-1, it was found that sequences
of radial velocities showed deviations that were correlated over
periods of about one day (Van Kerkwijk et al. \cite{VK95}).  To see
whether such deviations are present in HD~153919 as well, in
Fig.~\ref{rvcurve} lines connect residual velocities for spectra that
were taken as part of a sequence.  Unfortunately, the result is
inconclusive.  Some groups seem to show correlated deviations, but for
others the deviations seem random.  It would be certainly worthwhile
to monitor HD~153919 during some nights continuously to have a clue to
the origin of those excursions and stellar wind variations that
certainly influence and hamper an accurate determination of the
orbital parameters.

Finally, we should mention that almost half of the set of spectra was
taken in the small aperture of the SWP camera, the others in the large
aperture (see Table~\ref{table1}). The spectra taken in the small
aperture are the ones of the first years of IUE and have
generally lower S/N.  We determined orbital solutions for the small
and large-aperture sets separately, but did not find significant
differences.

\section{Comparison with earlier results}

From the present results we can conclude that the derived radial
velocities are in good agreement with the results obtained much
earlier from optical data (see Hammerschlag-Hensberge \cite{HH78}).

Our results do not differ much from those by Heap \& Corcoran (\cite
{HC92}) for the same data. The advantage of our analysis is that we
use a larger part of the spectrum and correlate all spectra with each
other, which, as a consequence, results in a smaller error on
$K$. Remarkably, Heap \& Corcoran (\cite {HC92}) find a negligible
eccentricity ($e \sim 0.003$) with a very small error of the order of
$0.003$. At first glance, the radial velocities in their Fig. 2 seem
to fit equally well to an eccentric orbit. Also, from a statistical
point of view, their errors on $e$ and $\omega$ are much too
small. For small eccentricities the deviation from a circular orbit is
a sinusoid with half the binary period and it is easy to see that $e
\sim A(P/2)/A(P) = A(P/2)/K$, where $A$ is the amplitude of the
sinusoid and $K$ the radial velocity amplitude. The error on $A(P/2)$
will be comparable to the one on $K$ and as a consequence $\sigma_{e}
\sim \sigma_{K}/K$ .  This implies $\sigma_e \sim 0.17$ for the error
on the eccentricity, instead of $\sigma_e = 0.003$ found by Heap \&
Corcoran.

On the other hand, Stickland \& Lloyd (\cite {SL93}) find a stellar
orbit with an eccentricity close to ours, but their results show a
much smaller semi-amplitude $K$ of 10~\kms. Contact by one of the
authors with Dave Stickland for similar discrepant velocity values in
the X-ray binary Vela X-1 has proven that the main cause for the
discrepancy in the results is that their cross-correlation procedure
sets to zero the regions around interstellar lines and wind
lines. When this is combined with slight inadequacies in
normalization, artificial lines are introduced at fixed wavelengths in
both spectra to be cross-correlated. As a consequence the radial
velocities are biased systematically towards zero and the
radial-velocity amplitude will be underestimated. A discrepancy of
about 4~\kms\ had been found between the radial-velocity amplitude for
the ultraviolet spectra of Vela X-1 analyzed by both groups (see
Barziv et al. \cite{BK00} and Stickland et al. \cite{SL97}). The
reason for the much larger discrepancy of 10~\kms\ for HD153919 might
be the weaker and broader photospheric lines in this star combined
with the better signal-to-noise ratio which give more striking errors
when interstellar and wind lines are set to zero.

\section{Conclusions and evolutionary consequences}

The two main conclusions of this paper are that (i) the
radial-velocity amplitude $K$ of HD153919 determined from ultraviolet
spectra is consistent with that derived from optical spectra; and (ii)
the orbit most likely has a significant eccentricity. Note that the
obtained value for $K$ depends on whether the orbit is eccentric
($K=20.6$~\kms, $e=0.22$) or not ($K=18.7$~\kms).

Clark et al. (\cite {CG02}) performed a thorough study of the system
parameters of HD153919/4U1700-37 and determined the most likely masses
of both components using Monte Carlo simulations. These suggest a mass
for HD153919 and 4U1700-37 of ${\rm M}_{\star} = 58 \pm
11$~M$_{\odot}$ and ${\rm M}_{X} = 2.4 \pm 0.3$~M$_{\odot}$,
respectively. Clark et al. used the results described in this paper in
their analysis, but prefered the orbital solution with zero
eccentricity. As the mass function depends on the eccentricity:
\begin{equation} 
f = 1.038 \times 10^{-7} K^{3} P (1 - e^{2})^{3/2}
\end{equation} 
and 
\begin{equation} ({\rm M}_{\star} + {\rm M}_{X})^{2} = 
        \frac{ ({\rm M}_{X} \sin{i})^{3}}{f} \, , 
\end{equation}
they noticed that the masses of both components become significantly
higher when taking $e=0.22$. They argue that such high values for
$M_{\star}$ are inconsistent with $\log{g}$ derived from the analysis
of the photospheric spectrum. A non-zero eccentricity also results in
a higher number of solutions that are rejected due to the inclination
constraints. However, Clark et al. took the value of the
radial-velocity amplitude belonging to the eccentric orbit
($K=20.6$~\kms, so that $f = 0.0031 $~M$_{\odot}$), rather than $K=18.7$~\kms\ for
the circular orbit ($f = 0.0023 $~M$_{\odot}$); for the latter value the masses
would have become even higher. If one takes $K=20.6$~\kms\ and
$e=0.22$ then $f = 0.0029 $~M$_{\odot}$. Therefore, we conclude that the masses of
both components are about 4~\% higher than proposed by Clark et
al. (\cite{CG02}) if one uses the correct value for $f$ in case of the
eccentric orbital solution.

The short orbital period of the system, combined with its
eccentricity, will induce strong tidal interactions between the two
components. Quaintrell et al. (\cite{QN03}) present evidence for tidally
induced non-radial oscillations in HD77581, the B-supergiant companion
to Vela~X-1, like 4U1700-37 an eccentric ($e=0.09$), wind-fed high-mass
X-ray binary with an orbital period of 8.9 days. These non-radial
oscillations may well be the cause of the radial-velocity excursions
detected in Vela~X-1. The radial-velocity excursions in HD153919 could
have a similar origin.

The eccentricity of the orbit of both Vela~X-1 and 4U1700-37 suggests that
these systems have just entered the high-mass X-ray binary phase and
are on their way to become Roche-lobe overflow systems. The latter
systems have circular orbits, possibly the result of tidal
interaction. Both HD77581 and HD153919 do not rotate synchronously
with their orbit: $P_{\rm orb} / P_{\rm rot}$ is 0.67 (Zuiderwijk
1995) and 0.46 (taking $v \sin{i} = 150$~\kms, $i=90^{\circ}$,
$R_{\star} = 21.9$~R$_{\odot}$), respectively, which may indicate that
the supergiants are still expanding and that tidal forces have not
(yet?) managed to synchronize the system and circularize the
orbit. 

Conti (\cite{Co78}) and Rappaport \& Joss (\cite{RJ83}) suggested that the OB
supergiants in HMXBs are too luminous for their mass. Kaper (\cite{Ka01})
showed that the discrepancy between the measured and "spectroscopic"
mass is most severe for the Roche-lobe overflow systems, but is only
modest for wind-fed systems such as Vela~X-1. The proposed high mass
of HD153919 is consistent with its high luminosity, again suggesting
that HD153919 has not yet reached the phase of Roche-lobe overflow.

\begin{acknowledgements}
LK and MHvK acknowledge support from a fellowship of the Royal
Netherlands Academy of Arts and Sciences.
\end{acknowledgements}


\begin{thebibliography}{}
\bibitem[2001]{AK01} Ankay, A., Kaper, L., De Bruijne, J.H.J., et
al. 2001, A\&A 370, 170
\bibitem[2001]{BK00}
Barziv, O., Kaper, L., van Kerkwijk, M.H., Telting, J.H., \& van
Paradijs, J. 2001, A\&A 377, 925
\bibitem[2002]{CG02}
Clark, J.S., Goodwin, S.P., Crowther, P.A., et al. 2002, A\&A 392, 909
\bibitem[1978]{Co78}
Conti, P.S. 1978, A\&A 63, 255
\bibitem[1988]{Ev88}
Evans, N.R. 1988, IUE NASA Newsletter 17, 53
\bibitem[1986]{GB86}
Gies, D.R. \& Bolton, C.T. 1986, ApJS 61, 419
\bibitem[1989]{HW89}
Haberl, F., White, N.E., \& Kallman, T.R. 1989, ApJ 343, 409
\bibitem[1978]{HH78}
Hammerschlag-Hensberge, G. 1978, A\&A 64, 399
\bibitem[1990]{HHK90} Hammerschlag-Hensberge, G., Howarth, I.D., \&
Kallman, T.R. 1990, ApJ 352, 698
\bibitem[1992]{HC92}
Heap, S.R., \& Corcoran, M.F. 1992, ApJ 387, 340
\bibitem[1973]{Jo73}
Jones, C., Forman, W., Tananbaum, H., et al. 1973, ApJ 181, L43
\bibitem[1998]{Ka98}
Kaper, L. 1998, in Boulder-Munich II: Properties of Hot, Luminous
Stars, ASP Conference Series Vol. 131, ed. Ian D. Howarth
(Astron. Soc. of the Pacific, San Francisco), 427
\bibitem[2001]{Ka01} Kaper, L. 2001, in The influence of binaries on
stellar population studies, Astr. \& Space Science Lib.,
ed. D. Vanbeveren (Kluwer Acad. Publ.), 125
\bibitem[1990]{KH90} Kaper, L., Hammerschlag-Hensberge, G., \& Takens,
R.J. 1990, Nature 347, 652
\bibitem[1994]{KH94} Kaper, L., Hammerschlag-Hensberge, G., \&
Zuiderwijk, E.J. 1994, A\&A 289, 846
\bibitem[1978]{Hu78}
Humphreys, R.M. 1978, ApJS, 38, 309
\bibitem[2003]{QN03} Quaintrell, H., Norton, A.J., Ash, T.D.C. et
al. 2003, A\&A in press (astro-ph/0301243)
\bibitem[1983]{RJ83}
Rappaport, S.A. \& Joss, P.C. 1983, in Accretion Driven Stellar X-ray
Sources, ed. W.H.G. Lewin \& E.P.J. van den Heuvel (Cambridge
University Press), 1
\bibitem[1996]{RF96}
Rubin, B.C., Finger, M.H., Harmon, B.A., et al. 1996, ApJ 459, 259
\bibitem[1992]{St92}
Stickland, D.J. 1992,in Complementary Approaches to Double and Multiple Star
Research, ASP Conference Series Vol. 32, ed. McAlister \& Hartkopf
(Astron. Soc. of the Pacific, San Francisco), 393
\bibitem[1993]{SL93}
Stickland, D.J., \& Lloyd, C. 1993, MNRAS 264, 935
\bibitem[1997]{SL97}
Stickland, D.J., Lloyd, C., \& Radzium-Woodham, A. 1997, MNRAS 286, L21
\bibitem[1993]{VK93}
Van Kerkwijk, M.H. 1993, PhD thesis, University of Amsterdam
\bibitem[1995]{vK95}
Van Kerkwijk, M.H., Van Paradijs, J., \& Zuiderwijk, E.J. 1995a, A\&A
303, 497
\bibitem[1995]{VK95} Van Kerkwijk, M.H., Van Paradijs, J., Zuiderwijk,
E.J., et al. 1995b, A\&A 303, 483
\bibitem[1977]{vP77}
Van Paradijs, J., Takens, R., \& Zuiderwijk, E.J. 1977, A\&A 57, 221
\bibitem[1995]{Zu95}
Zuiderwijk, E.J. 1995, A\&A 299, 79

\end{thebibliography}
\end{document}